\documentstyle[12pt]{article}
\renewcommand {\c}  {\'{c}}

\newcommand   {\cc} {\v{c}}

\renewcommand {\d}  {\hbox{d\hskip-1.1ex{\raise0.640ex\hbox{--}}\skip 0.70ex}}
\newcommand   {\D}  {\hbox{D\hskip-1.9ex{\raise0.175ex\hbox{--}}\hskip0.85ex}}
\newcommand   {\s}  {\v{s}}

\begin{document}

\renewcommand{\thesubsection}{\arabic{subsection}.}
\newcounter{buco}
\setcounter{equation}{0}
\setcounter{buco}{0}
\renewcommand{\theequation}{\arabic{subsection}.\arabic{equation}}

\begin{center}
\section*{Covariant - tensor method for quantum groups and applications I :
$SU(2)_{q}$}

\vspace*{20mm}

                      {\bf S. Meljanac}\\

                Rudjer Bo\s kovi\c \ Institute,\\
               Bijeni\cc ka c.54, 41001 Zagreb,\\
                          Croatia

\vspace*{10mm}

                       {\bf M. Milekovi\c }\\

           Prirodoslovno - Matemati\cc ki fakultet,\\
            Department of Theoretical Physics,\\
              Bijeni\cc ka c. 32, 41000 Zagreb,\\
                          Croatia

\vspace*{30mm}

Short title: Covariant - tensor method for quantum groups\\

Classification number: 02.20.+b
\end{center}
\newpage

\subsection*{Abstract}
\setcounter{equation}{0}

A covariant - tensor method for $SU(2)_{q}$  is described.  This
tensor method is used to calculate q - deformed Clebsch  -
Gordan  coefficients.  The   connection   with   covariant
oscillators   and   irreducible   tensor   operators    is
established.  This  approach  can  be  extended  to  other
quantum groups.

\newpage
\subsection{Introduction}
\setcounter{equation}{0}

In recent years there has been  considerable  interest
in q-deformations of Lie algebras (quantum groups) [1] and
their applications in physics [2]. The main goal of  these
applications  is  a  generalization  of  the  concept   of
symmetry. The properties of quantum groups are similar  to
those of classical Lie groups with q not being a  root  of
unity. However, it is still not clear to what  extent  the
familiar tensor methods, used in the representation theory
of Lie algebras,  are  applicable  to  the  case  of  q  -
deformations.

    Different types of the tensor calculus for $ SU(2)_{q} $  were
proposed and applied in references [3,4,6,9,10].  However,
no simple covariant  -  tensor  calculus  for $ SU(n)_{q}$   was
presented. In this paper we propose a simple  covariant  -
tensor method for$ SU(2)_{q}$  which  can  be  extended  to  the
general $ SU(n)_{q}$ . Details  for $ SU(n)_{q} $   and  especially  for
$ SU(3)_{q}$  will be published separately.

    The plan of the paper is the following. In  Section  2
we  recall  the  basics  of  the $  SU(2)_{q} $    algebra,   its
fundamental representation and invariants. In Section 3 we
construct the  general  $ SU(2)_{q} $   -  covariant  tensors  and
invariants. In Section 4 we apply this  tensor  method  to
calculate q - deformed Clebsch-Gordan coefficients and  in
Section 5 we demonstrate their symmetries.  We  point  out
that this method is simpler than  that  used  in  previous
calculations [5,6] and can be generalized to other quantum
groups. Finally, in Section 6 we connect covariant tensors
with  covariant  q  -  oscillators  and   construct   unit
irreducible tensor operators.

\subsection{$ SU(2)_{q}$  - algebra, its fundamental  representation  and
invariants}
\setcounter{equation}{0}

    Let us recall that three generators of $ SU(2)_{q}$  obey the
following commutation relations [1] (we take q real)

\begin{displaymath}
[J^{0}, J^{\pm}] = \pm J^{\pm}
\end{displaymath}

\begin{equation}
[J^{+}, J^{-}] = [2J^{0}]_{q} = \frac {q^{2J^{0}}- q^{-2J^{0}}} {q-q^{-1}}
\end{equation}

The coproduct $\Delta$ \ ;\  $ SU(2)_{q}\rightarrow  SU(2)_{q}\otimes
SU(2)_{q}$  is defined as

\begin{equation}
\begin{array}{c}
\Delta (J^{\pm}) = J^{\pm} \otimes q^{J^{0}} + q^{-J^{0}}\otimes J^{\pm}\\
\Delta (J^{0}) = J^{0} \otimes 1 + 1 \otimes J^{0}
\end{array}
\end{equation}

Let $ V_{2} $  be a two - dimensional  space  spanned  by  the
basis $|e_{a} >, a=1,2$ , and $|v > = \sum_{a} | e_{a} >v_{a} \ \in \ V_{2} $ .
 The
$ SU(2)_{q} $  generators $ J^{k}  (k= \pm,o)$ act as

\begin{displaymath}
J^{k}|e_{a}\rangle = \sum_{b}(J^{k})_{ba}|e_{b}\rangle
\end{displaymath}

\begin{equation}
J^{k}|v\rangle = \sum_{a,b} (J^{k})_{ba}|e_{b}\rangle v_{a}
= \sum_{b} |e_{b}\rangle (J^{k}v)_{b}
= \sum_{b} |e_{b}\rangle v^{,}_{b}.
\end{equation}

In  the   fundamental   representation   of $  SU(2)_{q}$    the
generators $ J^{k}s $ are ordinary 2x2 Pauli matrices.
    Let $(V_{2})^{\ast}$  be a  dual  space  with  the  basis
$<e_{a}| = (|e_{a} >)^{+}$  and $<v|= (|v>)^{+}  =\sum_{a}
v^{\ast}_{a}<e_{a}|.$
The  dual  basis is orthonormal, i.e. $<e_{a} | e_{b} >  = \delta_{ab}$.
We  note  that  the components of the vector $|v>, v_{a}$, ( or $v^{\ast}_{a}$
of $<v|$) are not
defined  as  real  or  complex  numbers.  Their  algebraic
properties    will    follow    from  $SU(2)_{q}$ -invariance
requirements. Here we identify ( for the spin $j = 1/2$ )

\begin{equation}
\begin{array}{c}
|e_{a}\rangle = |\frac {1}{2}, m_{a} \rangle\\
\langle e_{a}| = \langle \frac {1}{2}m_{a}|\\
m_{a} = \pm \frac {1}{2}
\end{array}
\end{equation}

and the matrix elements of the generators $J_{k}$  are

\begin{equation}
\begin{array}{c}
\langle e_{a}|J^{0}|e_{a}\rangle = m_{a}\\
\langle e_{1}|J^{+}|e_{2}\rangle = \langle e_{2}|J^{-}|e_{1}\rangle = 1
\end{array}
\end{equation}

We define a scalar product $as <u|v> = \sum_{a} u^{\ast}_{a}v_{a}$  and the
norm
as \linebreak[4] $ <v|v>= \sum_{a} v^{\ast}_{a} v_{a}$. This scalar product
(and  the  norm)
are not $SU(2)_{q}$  - invariant. Instead, the quantity

\begin{equation}
\langle v|q^{-J^{0}} |v\rangle
\end{equation}

is invariant under the action of the  coproduct  (2.2)  in
the following sense:

\begin{equation}
\Delta (J^{\pm})\langle v|q^{-J^{0}}|v\rangle=(J^{\pm}\langle v|)|v\rangle +
(q^{-J^{0}}
\langle v|) J^{\pm} q^{-J^{0}}|v\rangle
= -\langle v| J^{\pm}|v\rangle + \langle v|J^{\pm}|v\rangle = 0,
\end{equation}

\begin{equation}
\Delta(J^{0})\langle v|q^{-J^{0}} |v\rangle = (J^{0}\langle
v|)q^{-J^{0}}|v\rangle + \langle v|J^{0}q^{-J^{0}} |v\rangle
= - \langle v|J^{0}q^{-J^{0}}|v\rangle + \langle v|J^{0} q^{-J^{0}}|v\rangle
= 0.
\end{equation}

The quadratic forms
\begin{equation}
\sum_{a} u^{\ast}_{a} q^{-J^{0}} v_{a} = \sum_{a} u^{\ast}_{a} q^{-m_{a}}
v_{a}
\end{equation}

and

\begin{equation}
\sum_{a} v_{a} q^{m_{a}} u^{\ast}_{a}
\end{equation}

are $SU(2)_{q}$ -invariant. Note that the first  quadratic  form
(2.8) can be written as $ <u|q^{-J_{0}}|v>$.
If we demand $ \sum_{a} v^{\ast}_{a} q^{-m_{a}} v_{a}=\sum_{a} v_{a}q^{m_{a}}
v^{\ast}_{a}$, it  follows  that $ v^{\ast}_{1}v_{1}=q v_{1} v^{\ast}_{1}$ and
$ v^{\ast}_{2}v_{2}=q^{-1}v_{2}v^{\ast}_{2}.$

    In addition to the $ <u|q^{-J_{0}}|v>$ -  invariant  form  we
consider another form,

\begin{equation}
\epsilon_{ab}|e_{a} \rangle | e_{b}\rangle
\end{equation}

with

\begin{equation}
\begin{array}{c}
\epsilon_{ab} =
\left (
\begin{array}{cc}
0&q^{\frac{1}{2}}\\
-q^{-\frac{1}{2}}&0\\
\end{array}
\right )\\
\epsilon_{ab} \epsilon_{bc}=-\delta_{ac}\\
(\epsilon_{\overline {a}\overline {b}})_{q}=-(\epsilon_{a b})_{q^{-1}}
\end{array}
\end{equation}

where $\overline {1} = 2$ and $\overline {2} = 1$ .
Note that the q -  antisymmetric  combination $ v_{a} v_{b}\varepsilon_{ab}$ is
$SU(2)_{q}$ -invariant, showing that $v_{a}$  and $v_{b}$  do  not  commute.
Instead, they q - commute, i.e. $v_{2} v_{1}  = q v_{1} v_{2}$ .

\subsection{General $SU(2)_{q}$  - tensors and invariants}
\setcounter{equation}{0}

    Let  us  consider   the   tensor   -   product space $ (V_{2})^{\otimes
k}=V_{2}\otimes ...\otimes V_{2}$ with the basis \linebreak[4]
$|e_{a_{1}}>\otimes ...\otimes|e_{a_{k}}>, a_{1},...a_{k}=1,2$.
Then we write an element of the tensor space $ (V_{2})^{\otimes k}$ as tensor
$|T>$ of the form

\begin{equation}
\mid T\rangle= \mid e_{a_{1}}\rangle....\mid e_{a_{k}}\rangle T^{a_{1}}....
T^{a_{k}}
=\mid e_{a_{1}}...e_{a_{k}}\rangle T^{{a}_{1}...{a}_{k}}
\end{equation}

We have the following proposition:

  {\em The tensor $|T>$ transforms under the $SU(2)_{q}$  algebra as an
irreducible representation of spin $j=k/2$ if  and  only  if
$T^{2}T^{1}= q T^{1} T^{2}$.}

    Let us assume $T^{2}T^{1}  = q T^{1}T^{2}$ . Then

\begin{equation}
\mid T_{j=\frac {k}{2}}\rangle=\mid e_{a_{1}}... e_{a_{k}}\rangle T^{a_{1}...
a_{k}}
= \sum^{+j}_{m=-j} \mid jm\rangle T^{jm}
\end{equation}

The vectors $ |jm>$ span the space $V_{2j+1}$ of the irreducible
representation with spin  $j$.  From  $T^{2}T^{1}=qT^{1}T^{2}$ it
follows that

\begin{equation}
T^{a_{1}...a_{k}}=q^{\chi(a_{1},...a_{k}}) : T^{a_{1}...a_{k}}:
\end{equation}

where $:T:$ means the normal order of indices  (1's  on  the
left of 2's), i.e. $T^{11..122..2} $ and index 1 (2) appears  $n_{1}
(n_{2})$ times, respectively. $\chi$ is the  number  of  inversions
with respect to the normal order. Hence,

\begin{equation}
\mid jm \rangle = \mid e_{\{a_{1}...a_{k}\}} \rangle
= \frac {1}{\sqrt {f}} q^{-\frac {M}{2}} \sum_{perm(a_{1}...a_{k})}
q^{\chi(a_{1}...a_{k})}\mid e_{a_{1}...a_{k}}\rangle
\end{equation}

where   the   curly   bracket   $\{a_{1} ..a_{k}\}$ denotes  the
q-symmetrization. The summation runs over all the  allowed
permutations of the fixed set of indices $(n_{1}$ 1's  and $n_{2}$
2's) and

\begin{displaymath}
M=n_{1}n_{2}=(j+m)(j-m)
\end{displaymath}

\begin{equation}
j=\frac {1}{2} (n_{1}+n_{2}) \ \ \ \ \  m=\frac {1}{2} (n_{1}-n_{2})
\end{equation}

\begin{displaymath}
f=
\left(
\begin{array}{c}
2j\\
j+m
\end{array}
\right )_{q}
= \frac {\left[2j\right]_{q}!} {\left[j+m\right]_{q}!\left[j-m\right]_{q}!}
\end{displaymath}

The important relation is
\begin{equation}
f=q^{-M}\sum_{perm(a_{1}...a_{k})}q^{2\chi(a_{1}...a_{k})}
\end{equation}

{}From equation (3.4) and the definition  of  the  coproduct
$\Delta(J^{\pm})$ (2.2) we can reproduce

\begin{equation}
\begin{array}{c}
\Delta(J^{\pm})\mid jm\rangle=\sqrt {[j\mp m]_{q}[j\pm m+1]_{q}}
\mid j m \pm 1\rangle\\
\Delta(J^{0})\mid jm\rangle = m \mid jm\rangle
\end{array}
\end{equation}

{}From (3.2) and (3.4) we immediately  obtain  the  relation
between $T^{jm}$ and the components of $T^{a_{1}..a_{k}}$      :

\begin{equation}
\begin{array}{c}
T^{jm}=q^{\frac {M}{2}}\sqrt {f} : T^{a_{1}... a_{k}}:\\
T^{j-m}=q^{\frac {M}{2}} \sqrt {f} : T^{\overline {a}_{1}...\overline {a}_{k}}:
\end{array}
\end{equation}

where $\overline {1} = 2,\overline {2} = 1$ and $T^{j-m} =
(T^{jm})_{n_{1}\Leftrightarrow n_{2}}$.
    In the dual space $(V^{\otimes k}_{2})^{\ast}$ we define

\begin{equation}
\begin{array}{c}
\langle e_{a_{k}...a_{1}} \mid = (\mid e_{a_{1}...a_{k}}\rangle)^{+}\\
\langle e_{a_{k}...a_{1}} \mid e_{b_{1}...b_{k}}\rangle = \delta_{a_{1}b_{1}}
...\delta_{a_{k}b_{k}}
\end{array}
\end{equation}

and in the dual space $(V_{2j+1})^{\ast}$  we define

\begin{displaymath}
\langle jm\mid = (\mid jm\rangle)^{+} = \langle e_{\{a_{k}...a_{1}\}}\mid=
\end{displaymath}

\begin{equation}
= \frac {1}{\sqrt {f}} q^{-\frac {M}{2}} \sum _{perm(a_{1}...a_{k})}
q^{\chi(a_{1}...a_{k})} (\mid e_{a_{1}...a_{k}}\rangle)^{+}=
\end{equation}

\begin{displaymath}
= \frac {1}{\sqrt {f}} q^{-\frac {M}{2}} \sum_{perm(a_{1}...a_{k})}
q^{\chi(a_{1}...a_{k})} \langle e_{a_{k}...a_{1}}\mid.
\end{displaymath}

As a consequence of equations (3.4),(3.6)  and  ((3.9)  we
obtain

\begin{equation}
\langle jm_{1} \mid jm_{2}\rangle = \frac {1}{f} q^{-M} \sum_{perm(a_{1}...
a_{k})} q^{2\chi(a_{1}...a_{k})} \delta_{m_{1}m_{2}}
=\delta_{m_{1}m_{2}}.
\end{equation}

    The $SU(2)_{q}$ - invariant quantity built up of the tensors
$<T|$ and $|U>$ of spin $j = k/2$ is

\begin{equation}
I=\langle T\mid q^{-J^{0}}\mid U\rangle = (T^{a_{k}...a_{1}})^{\ast}
q^{-J^{0}}U^{a_{1}...a_{k}}
= \sum^{+j}_{m=-j} (T^{jm})^{\ast} U^{jm} q^{-m}.
\end{equation}

The second type of the $SU(2)_{q}$  - invariant  quantity  built
up of the tensors $|T>$ and $|U>$ of spin $j =k/2$ is

\begin{equation}
I^{'} = T^{a_{k}...a_{1}} U^{b_{1}...b_{k}} \epsilon_{a_{1}b_{1}}
\epsilon_{a_{2}b_{2}}...\epsilon_{a_{k} b_{k}}
\end{equation}

with $\epsilon_{ab}$ given in (2.11). Of course, $T^{a}$  $T^{b}$
$\epsilon_{ab} = 0$  if  $T^{a}$
and $T^{b}$  $q$-commute.
Furthermore, using equation (3.3) we can also write

\begin{equation}
\begin{array}{c}
I=q^{\chi(s)} (T^{a_{k}...a_{1}})^{\ast} q^{-J^{0}} U^{s(a_{1}...a_{k})}\\
I^{'}=q^{\chi(s)} T^{a_{k}...a_{1}} U^{s(b_{1}...b_{k})}\epsilon_{a_{1}b_{1}}
... \epsilon_{a_{k}b_{k}}
\end{array}
\end{equation}

where $s \in S_{k}$  is a fixed permutation of the indices $a_{1} ...a_{k}$
and $\chi (s) = \chi (a_{1} ...a_{k}) - \chi (s(a_{1} ...a_{k}))$  is  the
number  of
inversions with respect to the $(a_{1} ...a_{k})$ order.

\subsection{q - Clebsch-Gordan coefficients}
\setcounter{equation}{0}

    Here we present a new simple  method  for  calculating
the q-deformed  Clebsch-Gordan  coefficients.  It  can  be
immediately extended  and  applied  to  $ SU(n)_{q} $   and  other
quantum groups.  This  method  is  a  consequence  of  the
previously described tensor  method  and  construction  of
invariants.
    Our notation is

\begin{equation}
\mid JM\rangle = \sum_{m_{1},m_{2}} \langle j_{1}m_{1}j_{2}m_{2}\mid
JM\rangle_{q} \mid j_{1}m_{1}\rangle \mid j_{2}m_{2}\rangle.
\end{equation}

For $q \in R$, $C-G$ coefficients are real

\begin{equation}
\langle j_{1}m_{1} j_{2}m_{2} \mid JM\rangle^{\ast}_{q} = \langle j_{1}m_{1}
j_{2} m_{2}\mid JM\rangle _{q}
\end{equation}

and

\begin{equation}
\langle j_{1}m_{1}j_{2}m_{2}\mid JM\rangle_{q} = \langle JM\mid j_{1}
m_{1} j_{2}m_{2}\rangle_{q}
\end{equation}

Using the tensor notation $|jm\rangle = |e_{\{a_{1}...a_{k}\}}\rangle$ ((3.4),
(3.9)
and (3.10) ),  we  first  calculate  $C-G$  coefficient  for
$j_{1}\otimes j_{2}\rightarrow j_{1}+j_{2}$:

\begin{displaymath}
\langle j_{1}+j_{2}\ \ \  m_{1}+m_{2} \mid j_{1}m_{1}\ \ \  j_{2}m_{2} \rangle
_{q}=
\end{displaymath}

\begin{displaymath}
=\langle e_{\{b_{l}...b_{1},a_{k}...a_{1}\}}\mid e_{\{a_{1}...a_{k}\}}
e_{\{b_{1}...b_{l}\}}\rangle=
\end{displaymath}

\begin{displaymath}
= \langle e_{\{b,a\}}\mid e_{\{a\}}e_{\{b\}}\rangle=
\end{displaymath}

\begin{equation}
= \frac {1} { \sqrt {f_{1}f_{2}f_{3}} } q^{-\frac {1}{2} (M_{1}+M_{2}+M_{3})}
\sum_{perm(a),(b)} q^{\chi(a)+\chi(b)+\chi(a,b)}=
\end{equation}

\begin{displaymath}
= \sqrt { \frac {f_{1}f_{2}} {f_{3}} } q^{\frac {1}{2}(M_{1}+M_{2}-M_{3})}
q^{(j_{1}-m_{1}) (j_{2}+m_{2})}=
\end{displaymath}

\begin{displaymath}
= \sqrt {\frac {f_{1}f_{2}} {f_{3}} }
q^{j_{1}m_{2}-j_{2}m_{1}}.
\end{displaymath}

where we have used

\begin{equation}
\chi(a,b)=\chi(a)+\chi(b)+(j_{1}-m_{1}) (j_{2}+m_{2})
\end{equation}

and equation (3.6) together with the abbreviations

\begin{displaymath}
k=2j_{1}\ \ \ l=2j_{2}\ \ \ j_{3}=j_{1}+j_{2}\ \ \ m_{3}=m_{1}+m_{2}
\end{displaymath}

\begin{displaymath}
M_{i}=(j_{i}+m_{i}) (j_{i}-m_{i})
\end{displaymath}

\begin{equation}
f_{i}=\left(
\begin{array}{c}
2j_{i}\\
j_{i}+m_{i}
\end{array}
\right )_{q}
\end{equation}

\begin{displaymath}
\frac {f_{1}\cdot f_{2}} {f_{3}} =
\frac
{[2j_{1}]_{q}![2j_{2}]_{q}![j_{3}+m_{3}]_{q}! [j_{3}-m_{3}]_{q}!}
{[j_{1}+m_{1}]_{q}![j_{1}-m_{1}]_{q}! [j_{2}+m_{2}]_{q}!
[j_{2}-m_{2}]_{q}![2j_{3}]_{q}!}
\end{displaymath}

The main  observation  is  that  any  $C-G$  coefficient
$\langle j_{1} m_{1}j_{2}m_{2} |JM\rangle$ can be written in the form (4.4)  .
Namely,
the $C.-G.$ coefficient $\langle j_{1} m_{1} j_{2} m_{2}|JM\rangle$ is
projection of  the
state $\langle j_{1} m_{1}|\otimes \langle j_{2}m_{2}| = \langle
e_{\{a_{1}...a_{2j_{1}}\}} e_{\{b_{1}...b_{2j_{2}}\}}|$
from the tensor product space $V^{\ast}_{2j_{1}+1}\otimes V^{\ast}_{2j_{2}+1}$
to the  state $|JM\rangle    =       |
e_{\{a_{1}..[a_{2j_{1}-n+1}[..[a_{2j_{1}},b_{1}]..]b_{n}]b_{n+1}..b_{2j_{2}}\}}\rangle$
(with the appropriate symmetry of $2j_{1} +2j_{2}$ indices) in  the
space $V_{2J+1}\subset  V_{2j_{1}+1}\otimes V_{2j_{2}+1}$.
Here, the  square  brackets $[..]$ denote $q$-antisymmetrization and $n = 2j =
j_{1} +j_{2} -J$.

Furthermore,  the  state  $|
e_{[a_{1}[a_{2}..[a_{n},b_{n}]..b_{2}]b_{1}]}>\infty \
\epsilon_{a_{n}b_{n}}...\epsilon_{a_{1}b_{1}}$  transforms  as  a  singlet,
i.e.  it
is invariant under the coproduct action in the tensor product
space $V_{n}\otimes V_{n}$. Hence, using the  equation  (3.4),  we  can
write

\begin{displaymath}
\langle j_{1}m_{1}j_{2}m_{2}|JM\rangle _{q}={\cal N}{\sum_{perm(a,b),(c,d)}}
\langle e_{\{a,b\}} e_{\{c,d\}}|e_{\{a,d\}}\rangle \cdot (\epsilon_{(b,c)})_{n}
\end{displaymath}

\begin{equation}
= {\cal N}
\frac {q^{-\frac {1}{2}(M_{1}+M_{2}+M_{J})}} {\sqrt {f_{1}\cdot f_{2}\cdot
f_{J}}}
{\sum_{perm (a,b),(c,d)}} q^{\chi(a,b)+\chi(c,d)+\chi(a,d)}
(\epsilon_{(b,c)})_{n}
\end{equation}

where the length of $b (c)$ is $n = j_{1}+ j_{2}  - J, (\epsilon_{(b,c)})_{n}
= \epsilon_{b_{1}c_{1}}...\epsilon_{b_{n}c_{n}}$  and

\begin{equation}
{\cal N}=\left (
\frac {[2j_{1}]_{q}![2j_{2}]_{q}! [2J+1]_{q}}
{[j_{1}+j_{2}-J]_{q}![j_{1}-j_{2}+J]_{q}!
[-j_{1}+j_{2}+J]_{q}]![j_{1}+j_{2}+J+1]_{q}!}\right )^{\frac {1}{2}}.
\end{equation}

Expression (4.7) is efficient for practical calculation of
$C-G$ coefficients (see Appendix).
    We also present a simple derivation  of  the  standard
expression for $q - C-G$ coefficients [6]
Using the decomposition

\begin{displaymath}
<j_{1}m_{1}|=\sum^{+j}_{m=-j}<j_{1}m_{1}|j_{1}-j \ m_{1}-m \ j m>_{q}<j_{1}-j \
m_{1}-m|<jm|
\end{displaymath}

\begin{equation}
<j_{2}m_{2}|=\sum^{+j}_{m=-j}<j_{2}m_{2}|j-m \ j_{2}-j \
m_{2}+m>_{q}<j-m|<j_{2}-j \ m_{2}+m|
\end{equation}

\begin{displaymath}
|JM>=\sum^{+j}_{m=-j}<j_{1}-j \ m_{1}-m \ j_{2}-j \ m_{2}+m|JM>_{q}|j_{1}-j \
m_{1}-m>|j_{2}-j \ m_{2}+m>
\end{displaymath}

we immediately write

\begin{displaymath}
<j_{1}m_{1}j_{2}m_{2}|JM>_{q}=N \sum^{+j}_{m=-j} <j_{1}m_{1}|j_{1}-j \ m_{1}-m
\ jm>_{q}
\end{displaymath}

\begin{equation}
\times <j_{2}m_{2}|j-m \ j_{2}-j \ m_{2}+m>_{q}<jm \ j-m|00>_{q}
\end{equation}

\begin{displaymath}
\times <j_{1}-j \ m_{1}-m j_{2}-j \ m_{2}+m|JM>_{q}
\end{displaymath}

where N is the norm depending on $j_{1},j_{2}$  and $J$. Three of the
four $C-G$ coefficients appearing on the right  - hand  side
have  the  simple  form   (4.4). The  fourth   coefficient
$ <jm\ j-m|00>$ also has a simple form.  Namely,  for $n=2j$  we
have

\begin{displaymath}
<jm \ j-m|00>_{q}=\frac {1}{\sqrt {[n+1]_{q}}}
\epsilon_{a_{1}b_{1}}...\epsilon_{a_{n}b_{n}}=
\end{displaymath}

\begin{equation}
= \frac {1}{\sqrt {[2j+1]_{q}}} q^{\frac {1}{2} n_{1}} (-q^{{-\frac
{1}{2})}^{n_{2}}}=(-)^{j-m} \frac {1}{\sqrt {[2j+1]_{q}}} q^{m}.
\end{equation}

The denominator $\sqrt{[2j+1]}$  comes  from  the  orthonormality
condition.
    Finally, inserting  equations  (4.4)  and  (4.11) into
equation (4.10) we find

\begin{displaymath}
\langle j_{1}m_{1} \ j_{2}m_{2}|JM\rangle _{q}=N \sum^{+j}_{m=-j} \frac
{(-)^{j-m}} {\sqrt {[2j+1]_{q}}} q^{j_{1}m_{2}-j_{2}m_{1}}\times
\end{displaymath}

\begin{equation}
\times q^{m(2J+2j+1)}\frac {
\left (
\begin{array}{c}
2j\\
j+m
\end{array}
\right )_{q}
\left (
\begin{array}{c}
2j_{1}-2j\\
j_{1}-j+m_{1}-m
\end {array}
\right )_{q}
\left (
\begin{array}{c}
2j_{2}-2j\\
j_{2}-j+m_{2}+m
\end{array}
\right )_{q}}{
\sqrt {\left (
\begin{array}{c}
2J\\
J+M
\end{array}
\right )_{q}
\left (
\begin{array}{c}
2j_{1}\\
j_{1} +m_{1}
\end{array}
\right )_{q}
\left (
\begin{array}{c}
2j_{2}\\
j_{2} +m_{2}
\end{array}
\right )_{q}} }
\end{equation}

with $j_{1}+j_{2}-j=J+j$. This result  agrees  with  the  result found by Ruegg
[6] if the normalization factor $N$ is  taken as

\begin{equation}
N=\left \{ \frac
{[2j_{1}]_{q}![2j_{2}]_{q}![2J+1]_{q}![j_{1}+j_{2}-J+1]_{q}}
{[j_{1}+j_{2}-J]_{q}![j_{1}-j_{2}+J]_{q}![-j_{1}+j_{2}+J]_{q}![j_{1}+j_{2}+J+1]_{q}!}
\right \}^{\frac {1}{2}}
\end{equation}

We point out that our tensor method is simple and  can  be
easily applied to $SU(n)_{q}$  for $n\geq3$. We also mention that  it
can be  applied  to  multiparameter  quantum  groups.  For
example, it can be shown [7] that $C-G$ coefficients for the
two - parameter  $SU(2)_{p,q}$[8]  depend  effectively on one
parameter only.

\subsection{Symmetry relations}
\setcounter{equation}{0}

    For  completeness  we  rederive  the  known   symmetry
relations for $q - C-G$ coefficients and $q - 3-j$  symbols.
{}From equation (4.4) immediately follow symmetry relations

\begin{equation}
\begin{array}{c}
<j_{1}-m_{1} \ j_{2}-m_{2}|j_{1}+j_{2} \ -m_{1}-m_{2}>_{q}=\\
=<j_{2}m_{2} \ j_{1}m_{1}|j_{1}+j_{2} \ m_{1}+m_{2}>_{q}=\\
=<j_{1}m_{1} \ j_{2}m_{2}|j_{1}+j_{2} \ m_{1}+m_{2}>_{q^{-1}}
\end{array}
\end{equation}

and

\begin{displaymath}
<j_{1}-m_{1} \ j_{1}+j_{2} \ m_{1}+m_{2}|j_{2}m_{2}>_{q}=(-)^{j_{1}+m_{1}}
q^{-m_{1}}
\end{displaymath}

\begin{equation}
\times
\sqrt { \frac {[2j_{2}+1]_{q}} {[2j_{1}+2j_{2}+1]_{q}}}
<j_{1}m_{1} \ j_{2}m_{2}|j_{1}+j_{2} \ m_{1}+m_{2}>_{q}.
\end{equation}

Furthermore, from equation (4.11) we have
\begin{equation}
\begin{array}{c}
<j-m \ jm|00>_{q} = (-)^{2j} <jm \ j-m|00>_{q^{-1}}\\
<jm \ 00|jm>_{q} = 1.
\end{array}
\end{equation}

The  symmetry  relations  (5.1)-(5.3)  are  sufficient  to
derive the symmetries of  the  general  $C-G$  coefficients.
{}From equation (4.10) we obtain

\begin{equation}
\begin{array}{c}
<j_{1}-m_{1} \ j_{2}-m_{2}|J-M>_{q}
= <j_{2}m_{2} \ j_{1} m_{1}|JM>_{q}=\\
= (-)^{j_{1}+j_{2}-J} <j_{1}m_{1}\ j_{2}m_{2}|JM>_{q^{-1}}
\end{array}
\end{equation}

and

\begin{displaymath}
<j_{1}-m_{1}\ JM|j_{2}m_{2}>_{q}
= (-)^{J-j_{2}+m_{1}} q^{-m_{1}}
\end{displaymath}

\begin{equation}
\times
\sqrt {
\frac {[2j_{2}+1]_{q}} {[2J+1]_{q}}} <j_{1}m_{1}\ j_{2}m_{2}|JM>_{q}.
\end{equation}

(One can deduce this directly from (4.7) )

We can define the $q$-deformed $3-j$ symbol as

\begin{equation}
\left (
\begin {array} {c c c}
j_{1} &j_{2} &j_{3}\\
m_{1} &m_{2} &m_{3}
\end{array}
\right )_{q}
=q^{\frac {1}{3}(m_{2}-m_{1}) }
\frac {(-)^{j_{1}-j_{2}-m_{3}}} {\sqrt {[2j_{3}+1]_{q}}}
<j_{1}m_{1}\ j_{2}m_{2}|j_{3}-m_{3}>_{q}
\end{equation}

where the additional factor $q^{1/3(m_{2}-m_{1})}$ comes  from  the
requirement  that  symmetry  relations  for   the   $(3-j)_{q}$
coefficients should not contain explicit $q$-factors:

\begin{equation}
\left(
\begin{array} {c c c}
j_{1}& j_{2}& j_{3}\\
-m_{1}& -m_{2}& -m_{3}\\
\end{array}
\right)_{q}
=
\left(
\begin{array} {c c c}
j_{2}& j_{1}& j_{3}\\
m_{2}& m_{1}& m_{3}\\
\end{array}
\right)_{q}
=
(-)^{j_{1}+j_{2}+j_{3}}
\left(
\begin{array} {c c c}
j_{1}& j_{2}& j_{3}\\
m_{1}& m_{2}& m_{3}\\
\end{array}
\right)_{q^{-1}}
\end{equation}

and that  the  $(3-j)_{q}$ coefficients  are  invariant  under
cyclic permutations.
Note that the $ SU(2)_{q}$ invariant, built up  of  the  three
states $ |j_{1}m_{1}>, |j_{2}m_{2}>$ and $|j_{3}m_{3}>$, is

\begin{displaymath}
\sum_{m_{1},m_{2},m_{3}} <j_{3}-m_{3}\ j_{3}m_{3}|00>_{q} <j_{1}m_{1}\
j_{2}m_{2}|j_{3}-m_{3}>_{q}|j_{1}m_{1}>|j_{2}m_{2}>|j_{3}m_{3}>=
\end{displaymath}

\begin{equation}
=\sum_{m_{1},m_{2},m_{3}} q^{\frac {2}{3}(m_{1}-m_{3})}
\left(
\begin{array} {c c c}
j_{1}&j_{2}&j_{3}\\
m_{1}&m_{2}&m_{3}\\
\end{array}
\right)_{q}
|j_{1}m_{1}>|j_{2}m_{2}>|j_{3}m_{3}> =
\end{equation}

\begin{displaymath}
=\sum_{m_{1},m_{2},m_{3}} N_{123}
(\epsilon_{(b,c)})_{k_{1}}(\epsilon_{(d,e)})_{k_{2}}(\epsilon_{(a,f)})_{k_{3}}|e_{\{a,b\}}>|e_{\{c,d\}}>|e_{\{e,f\}}>.
\end{displaymath}

Now we identify

\begin{equation}
\begin{array}{c}
<j_{1}m_{1}\ j_{2}m_{2}|j_{3}-m_{3}>_{q} <j_{3}-m_{3}\ j_{3}m_{3}|00>_{q}=\\
=q^{\frac{2}{3}(m_{1}-m_{3})}
\left(\begin{array} {c c c}
j_{1}&j_{2}&j_{3}\\
m_{1}&m_{2}&m_{3}\\
\end{array}\right)_{q}=\\
=N_{123}(\epsilon_{(b,c)})_{k_{1}}(\epsilon_{(d,e)})_{k_{2}}(\epsilon_{(a,f)})_{k_{3}}
\end{array}
\end{equation}

where, for example,
$(\epsilon_{(b,c)})_{k}=\epsilon_{b_{1}c_{1}}...\epsilon_{b_{k}c_{k}}$ with

\begin{equation}
k_{1}=j_{1}+j_{2}-j_{3}\ \ \ k_{2}=-j_{1}+j_{2}+j_{3}\ \ \
k_{3}=j_{1}-j_{2}+j_{3}
\end{equation}

and $N_{123}$ is the normalization factor  fully  symmetric  in
indices (123). Equation (5.9)  represents  the  connection
with the tensor notation used.

\subsection{Covariant  q  -  oscillators  and irreducible  tensor
     operators}
\setcounter{equation}{0}

    Let us  define  the  q-bosonic  operators  $a_{i}$ and $a_{i}^{+}
(i=1,2)$ such that $ |e_{i}>=a_{i}^{+}|0,0>_{F}$ and $<e_{i}| =_{F}<0,0|
a_{i}$,
where $|0,0>_{F}$ denotes the  (Fock)  vacuum  state  invariant
under $SU(2)_{q}$ . Hence, $a^{+}_{1}$   and $a^{+}_{2}$  are  covariant
operators
transforming as an $SU(2)_{q}$  doublet. Therefore,  analogously
as in equation (3.2), they q-commute

\begin{equation}
a^{+}_{2} a^{+}_{1} = q \ a^{+}_{1} a^{+}_{2}.
\end{equation}

Furthermore, we define the  projector  $P_{(j=k/2)}$  from  the
tensor space  $(V_{2})^{\otimes k}$ to  the  totally  q-symmetric  space
carrying an irreducible representation of spin $j=k/2$

\begin{displaymath}
P_{(j=\frac {k}{2})}|e_{i_{1}...i_{k}}\rangle =\frac {1}{\sqrt {[k]_{q}!}}
a^{+}_{i_{1}}...a^{+}_{i_{k}}|0,0\rangle _{F}=
\end{displaymath}

\begin{equation}
=\frac {1}{\sqrt {[k]_{q}!}} q^{\chi(i_{1}...i_{k})} (a^{+}_{1})^{n_{1}}
(a^{+}_{2})^{n_{2}} |0,0\rangle _{F}.
\end{equation}

We find from equation (3.4) that

\begin{displaymath}
|jm>=q^{\frac {M}{2}} \frac {(a^{+}_{1})^{n_{1}} (a^{+}_{2})^{n_{2}}} {\sqrt
{[n_{1}]_{q}! [n_{2}]_{q}!}} |0,0\rangle _{F}
\end{displaymath}

\begin{equation}
j=\frac {1}{2} (n_{1}+n_{2})\ \ \  m=\frac {1}{2}(n_{1}-n_{2})
\end{equation}

We define the number operators $N_{i}$  and $N$ as

\begin{equation}
\begin{array}{c}
N_{i}|jm\rangle =N_{i}|n_{1},n_{2}\rangle =n_{i}|n_{1},n_{2}\rangle \\
N=N_{1}+N_{2}\ \ \ \left [N,N_{i}\right ] =0\ \ \ \left[ N_{1},N_{2}\right ]
=0\\
\left[ N_{i},a^{+}_{j}\right] = \delta _{ij}a^{+}_{i}\ \ \ \left[
N_{i},a_{j}\right ] =-\delta _{ij} a_{i}\\
\left[ N,a^{+}_{i}\right] =a^{+}_{i}\ \ \ \left[ N,a_{i}\right] =-a_{i}.
\end{array}
\end{equation}

The action of $a^{+}_{i}$  and $a_{i}$  on the basis vectors $|jm>$ is

\begin{equation}
\begin{array}{c}
a^{+}_{1}|jm>=q^{-\frac {1}{2} n_{2}}\sqrt {[n_{1}+1]_{q}}\ \ |j+\frac
{1}{2},m+\frac {1}{2}>\\
a^{+}_{2}|jm>=q^{ \frac {1}{2} n_{1}}\sqrt {[n_{2}+1]_{q}}\ \ |j+\frac
{1}{2},m-\frac {1}{2}>\\
a_{1}    |jm>=q^{-\frac {1}{2} n_{2}}\sqrt {[n_{1}]_{q}}  \ \ |j-\frac
{1}{2},m-\frac {1}{2}>\\
a_{2}    |jm>=q^{ \frac {1}{2} n_{1}}\sqrt {[n_{2}]_{q}}  \ \ |j-\frac
{1}{2},m+\frac {1}{2}>.
\end{array}
\end{equation}

The  commutation  relations  between  $a_{i}$   and  $a^{+}_{j}$ follow
immediately:

\begin{equation}
\begin{array}{c}
a^{+}_{2}a^{+}_{1}=q\ a^{+}_{1} a^{+}_{2}\ \ \ a_{2}a_{1}=q^{-1}\ a_{1}a_{2}\\
a_{2}a^{+}_{1}=a^{+}_{1}a_{2}\ \ \ a_{1}a^{+}_{2}=a^{+}_{2} a_{1}
\end{array}
\end{equation}

and

\begin{equation}
\begin{array}{c}
a_{1}a^{+}_{1}=q^{-N_{2}}\ [N_{1}+1]_{q}\ \ \ a^{+}_{1}a_{1}=q^{-N_{2}}\
[N_{1}]_{q}\\
a_{2}a^{+}_{2}=q^{+N_{1}}\ [N_{2}+1]_{q}\ \ \ a^{+}_{2}a_{2}=q^{+N_{1}}\
[N_{2}]_{q}\\
H=a^{+}_{1}a_{1}+a^{+}_{2}a_{2}=[N]_{q}
\end{array}
\end{equation}

Then
\begin{equation}
\begin{array}{c}
a_{1}a^{+}_{1}-q\ a^{+}_{1}a_{1}=q^{-N}\\
a_{2}a^{+}_{2}-q^{-1}\ a^{+}_{2}a_{2}=q^{+N}
\end{array}
\end{equation}

and

\begin{equation}
\begin{array}{c}
a_{1}a^{+}_{1}-q^{-1}\ a^{+}_{1}a_{1}=q^{2J^{0}}\\
a_{2} a^{+}_{2}-q\ a^{+}_{2} a_{2}=q^{2J^{o}}.
\end{array}
\end{equation}

The generators $J^{\pm}$  and $J^{0}$  can be represented as

\begin{equation}
\begin{array}{c}
J^{+}=q^{-J^{0}+\frac {1}{2}}\ a^{+}_{1} a_{2}\\
J^{-}=q^{-J^{o}-\frac {1}{2}} a^{+}_{2} a_{1}\\
2J^{0}=N_{1}-N_{2}\\
\left [ J^{+},J^{-}\right ]=[2J^{0}]_{q}=[N_{1}-N_{2}]_{q}\\
\left [N,J^{\pm}\right ]=[N,J^{0}]=0.
\end{array}
\end{equation}

We point out that the oscillator operators $a_{i}$  and $a^{+}_{i}$ are
covariant since  the  corresponding  tensors $|e_{\{i_{1}..i_{k}\}}>$,
equation  (3.4),  are   covariant   and   irreducible   by
construction.

    We note that the covariant q-Bose operators $a,a^{+}$  (6.1)
are the same as in [9], where they were constructed  using
the Wigner $D^{(j)}$-functions. A different  set  of  covariant
operators was constructed in ref.[10]. Other constructions
[11] are non-covariant in the sense that operators do  not
transform as $SU(2)_{q}$ doublet. In the non-covariant approach
one has to solve an  additional  problem  of  constructing
covariant, irreducible tensor operators [12].

The definition of the irreducible tensor operators  of
$SU(2)_{q}$ is

\begin{displaymath}
(J^{\pm}T_{km}-q^{-m}T_{km}J^{\pm})q^{-J^{o}}=
\end{displaymath}

\begin{displaymath}
=\sqrt {[k\mp m]_{q}[k\pm m+1]_{q}} \ T_{k\  m\pm 1}
\end{displaymath}

\begin{equation}
\left[J^{0}, T_{km}\right]=m\ T_{km}
\end{equation}

\begin{displaymath}
|jm>=T_{jm}|0,0>_{F}
\end{displaymath}

According to equations (6.1-6.3) we define a  unit  tensor
operator as

\begin{equation}
T_{jm}=
q^{\frac {1}{2} n_{1} n_{2}}
\frac
{(a^{+}_{1})^{n_{1}} (a^{+}_{2})^{n_{2}}}
{\sqrt {[n_{1}]_{q}! [n_{2}]_{q}!}}
\end{equation}

which is covariant and  irreducible  by  construction  and
satisfies the requirements (6.11) automatically.
Note that $(T_{km})^{+}$  transforms as contravariant tensor.
One can define the tensor

\begin{equation}
V_{k\mu}=(-)^{k-\mu}\ q^{\mu}\ T^{+}_{k-\mu}
\end{equation}

which transforms as covariant, irreducible tensor. In  the
tensor notation we have

\begin{equation}
V^{+}_{\{i_{1}...i_{k}\}}
=\epsilon_{i_{1}j_{1}}...\epsilon_{i_{k}j_{k}}
\ T_{\{j_{1}...j_{k}\}}=(-)^{n_{2}}
q^{\frac {1}{2}(n_{1}-n_{2})}\ T_{k-\mu}
\end{equation}

    For completeness, we  present  relations  between  the
Biedenharn operators $b_{i},b^{+}_{i}$  of ref.[11], $t_{i},t^{+}_{i}$  of
ref.[10]
and $a_{i}$,$a^{+}_{i}$  of the present paper:

\begin{equation}
\begin{array}{c}
b_{1}=q^{-N_{2}-\frac {1}{2}N_{1}}\ t_{1}=q^{\frac {1}{2}N_{2}} a_{1}\\
b_{2}=q^{-\frac {1}{2}N_{2}} t_{2}=q^{-\frac {1}{2}N_{1}} a_{2}\\
b^{+}_{1}=t^{+}_{1}q^{-N_{2}-\frac {1}{2}N_{1}}=a^{+}_{1}q^{\frac
{1}{2}N_{2}}\\
b^{+}_{2}=t^{+}_{2}q^{-\frac {1}{2}N_{2}}=a^{+}_{2}q^{-\frac {1}{2}N_{1}}
\end{array}
\end{equation}

We point out that the general covariant oscillators (e.g. $t_{i}, t^{+}_{i}$
and $a_{i}, a^{+}_{i})$ are characterized by the anyonic type q-commutation
relation (6.1).
Actually, equation (6.1) is a consequence of underlying braid group symmetry
and can be also obtained from the $ SU(2)_{q}$
R-matrix [10].

Finally,we give the Borel-Weil realization

\begin{equation}
a^{+}_{i}\equiv X_{i}\ \ \ a_{1}\equiv D_{i}\ \ \ i=1,2
\end{equation}

which  is   covariant   automatically.   The   commutation
relations are

\begin{equation}
\begin{array}{c}
X_{2}X_{1}=q\ X_{1}X_{2}\ \ \ D_{2}D_{1}=q^{-1}\ D_{1}D_{2}\\
D_{1}X_{1}=q\ X_{1}D_{1}+q^{-N}\ \ \ D_{2}X_{2}=q^{-1}X_{2}D_{2}+q^{N}\\
\left[D_{i},X_{j}\right]=0\ \ \ i\neq j
\end{array}
\end{equation}

or

\begin{equation}
\begin{array}{c}
D_{1}X_{1}=q^{-1}\ X_{1}D_{1}+q^{2J^{0}}\\
D_{2}X_{2}=q\ X_{2}D_{2}+q^{2J^{0}}
\end{array}
\end{equation}

where

\begin{equation}
\begin{array}{c}
N_{i}=X_{i}\partial_{i}\\
\partial_{i}=\frac {\partial}{\partial X_{i}}
\end{array}
\end{equation}

It follows that

\begin{equation}
\begin{array}{c}
D_{i} X^{n}_{i}=[n]_{q}\ X^{n-1}_{i}\\
D_{1}=\frac {1}{X_{1}}[X_{1}\partial_{1}]_{q}\ q^{-X_{2}\partial_{2}}\\
D_{2}=\frac {1}{X_{2}}[X_{2}\partial_{2}]_{q}\ q^{X_{1}\partial_{1}}
\end{array}
\end{equation}

\newpage

\section*{Appendix}
\setcounter{equation}{0}
\renewcommand{\theequation}{A.\arabic{equation}}

   We demonstrate usefullness of the  equation  (4.7)  for
the practical  calculations.  Using  equations  (4.5)  and
(4.11) we write :

\begin{equation}
\begin{array}{c}
\chi (a,b)=\chi (a)+\chi (b)+n_{2}(a)n_{1}(b)\\
\chi (c,d)=\chi (c)+\chi (d)+n_{2}(c)n_{1}(d)\\
\chi (a,d)=\chi (a)+\chi (d)+n_{2}(a)n_{1}(d)\\
\chi(b)=\chi(c)\\
(\epsilon_{(b,c)})_{n}=(-)^{n_{2}(b)} q^{\frac{1}{2}(n_{1}(b)-n_{2}(b))}
\end{array}
\end{equation}

where

\begin{equation}
\begin{array}{c}
n=2j=j_{1}+j_{2}-J\\
n_{1}(b)=n_{2}(c)=j+m\\
n_{2}(b)=n_{1}(c)=j-m\\
n_{1}(a)=j_{1}-j+m_{1}-m\\
n_{2}(a)=j_{1}-j-m_{1}+m\\
n_{1}(d)=j_{2}-j+m_{2}+m\\
n_{2}(d)=j_{2}-j-m_{2}-m\\
\end{array}
\end{equation}

After inserting equation (3.6)  into  equation  (4.7),  we
immediately obtain the final result,equation (4.12):

\begin{displaymath}
N
\frac {q^{-\frac {1}{2}(M_{1}+M_{2}+M_{J}) }}
{\sqrt {f_{1}\ f_{2}\ f_{J} }}
\sum^{2j}_{n_{1}(b)=0}
\sum_{perm(a)}
\sum_{perm(b)}
\sum_{perm(d)}
\end{displaymath}

\begin{displaymath}
\times q^{n_{2}(a)n_{1}(b)+n_{1}(b)n_{1}(d)+n_{2}(a)n_{1}(d)}
\end{displaymath}

\begin{displaymath}
\times q^{2\chi(a)+2\chi(b)+2\chi(d)}(\epsilon_(b,c))_{2j}=
\end{displaymath}

\begin{displaymath}
=N
\frac {q^{-\frac {1}{2}(M_{1}+M_{2}+M_{J}) }}
{\sqrt {f_{1}\ f_{2}\ f_{J} }}
\sum^{+j}_{m=-j}
q^{n_{2}(a)n_{1}(b)+n_{1}(b)n_{1}(d)+n_{2}(a)n_{1}(d)}
\end{displaymath}

\begin{equation}
\times f_{a}\ f_{b}\ f_{d}\
q^{n_{1}(a)n_{2}(a)+n_{1}(b)n_{2}(b)+n_{1}(d)n_{2}(d)} (\epsilon_{(b,c)})_{2j}
\end{equation}

\begin{displaymath}
=N
\sum^{+j}_{m=-j} (-)^{j-m}\ q^{j_{1}m_{2}-j_{2}m_{1}}\ q^{m(2J+2j+1)}
\end{displaymath}

\begin{displaymath}
\times \frac {f_{a}\ f_{b}\ f_{d}} {\sqrt {f_{1}\ f_{2}\ f_{J}}}.
\end{displaymath}

We extend this simple  calculation  of  the  $SU(2)_{q}$   C.-G.
coefficients  to  the  $SU(N)_{q}$   quantum   groups   in   the
forthcoming paper.

\newpage

\section*{Acknowledgements}
\setcounter{equation}{0}

This work was supported  by  the  joint  Croatian-American
contract NSF JF 999 and the Scientific Fund of Republic of
Croatia.

\end{document}